\documentclass[12pt]{article}
\usepackage{graphicx}
\usepackage{bm}
\def\Barcelo{Barcel\'o}

\begin{document}
\markright{Analogue models of and for gravity\hfil}
\title{\bf \LARGE
Analogue models of and for gravity}
\author{Matt Visser~$^*$, 
Carlos \Barcelo~$^\dagger$, and 
Stefano Liberati~$^\ddagger$
\\[2mm]  
{\small \it
\thanks{visser@kiwi.wustl.edu}
\thanks{carlos@hbar.wustl.edu; Now at Portsmouth University, England.}~Physics 
Department, Washington University,}
\\ {\small \it Saint Louis, Missouri 63130--4899, USA.}
\\
{\small \it
\thanks{liberati@physics.umd.edu}~Physics Department, University of Maryland}
\\
{\small \it College Park, Maryland 20742--4111, USA.}
}
\date{{
\small 26 November 2001; \LaTeX-ed \today }}
\maketitle
\begin{abstract}
\small
Condensed matter systems, such as acoustics in flowing fluids, light
in moving dielectrics, or quasiparticles in a moving superfluid, can
be used to mimic aspects of general relativity. More precisely these
systems (and others) provide experimentally accessible models of
curved-space quantum field theory.  As such they mimic
{\emph{kinematic}} aspects of general relativity, though typically
they do not mimic the {\emph{dynamics}}.  Although these analogue models
are thereby limited in their ability to duplicate all the effects of
Einstein gravity they nevertheless are extremely important --- they
provide black hole analogues (some of which have already been seen
experimentally) and lead to tests of basic principles of curved-space
quantum field theory.  Currently these tests are still in the realm of
{\emph{gedanken-experiments}}, but there are plausible candidate
models that should lead to laboratory experiments in the not too
distant future.
\\[5mm]
\end{abstract}

\clearpage 
\noindent 
{\emph{Plenary talk presented by Matt Visser. 
\\
Australasian Relativity Conference.
\\ 
Perth, Australia, July 2001.
\\
Proceedings to appear in General Relativity and Gravitation.}}
\def\g{{\mbox{\sl g}}}
\def\Box{\nabla^2}
\def\d{{\mathrm d}}
\def\half{{1\over2}}
\def\quarter{{1\over4}}
\def\L{{\cal L}}
\def\sech{\hbox{sech}}
\def\SIZE{1.00}
\def\ie{{\em i.e.\/}}
\def\eg{{\em e.g.\/}}
\def\etc{{\em etc.\/}}
\def\etal{{\em et al.\/}}
\def\Hospital{H\^opital}
\def\be{\begin{equation}}
\def\ee{\end{equation}}
\def\bea{\begin{eqnarray}}
\def\eea{\end{eqnarray}}
\section{Introduction}
\setcounter{equation}{0}

Analogue models of (and to some extent for) general relativity have
recently become a growth industry~\cite{Workshop}. Typically based on
various condensed-matter systems, these analogue models are most often
used for devising {\emph{gedanken-experiments}} that probe the
structure of curved-space quantum field theory. More boldly, they seem
promising routes to providing real laboratory tests of the foundations
of curved-space quantum field theory. (The most spectacular suggestion
along these lines is that analogue models may make experimental tests of
the Hawking radiation phenomenon a realistic possibility.)

Ideas along these lines have, to some extent, been quietly in
circulation almost since the inception of general relativity
itself. Walther Gordon (of the Klein--Gordon equation) introduced a
notion of ``effective metric'' to describe the effect of a refractive
index on the propagation of light~\cite{Gordon}. The Russian school,
as epitomized by Landau and Lifshitz, used notions developed in optics
to represent gravitational fields in terms of an ``equivalent
refractive index''~\cite{Landau-Lifshitz}.  There is an extensive, but
largely neglected {\emph{samizdat}} literature (of extremely variable
quality) that explores these issues. (For an extensive, though still
not comprehensive, bibliography see~\cite{Bibliography}.)

The modern revival is due largely to Unruh~\cite{Unruh} (and to some
extent Moncrief~\cite{Moncrief}) who in the early eighties considered
the use of hydrodynamic analogues, in which sound waves in a flowing
fluid are mapped into a suitable scalar field theory in an effective
curved spacetime --- the ``acoustic geometry''. (The precise
statement, as will be described more fully below, is that sound in an
irrotational inviscid barotropic fluid is identical to a massless
minimally coupled scalar field in curved spacetime; and
{\emph{quantized}} sound [the phonon field] is identical to
curved-space quantum field theory.)

The nineties saw considerable work on the nature of Hawking radiation
in these analogue models, still largely with the attitude that one was
performing {\emph{gedanken-experiments}}. It is only now, at the turn
of the millennium, that serious consideration is being given to the
actual construction of laboratory experiments. Three classes of system
stand out as being the most likely to lead to useful experimental
probes:
\begin{itemize}
\item
Acoustics in Bose--Einstein condensates.
\item
Slow light.
\item
Quasiparticles in superfluids.
\end{itemize}
%

\section{Acoustics in BECs}
\setcounter{equation}{0}

In this mini-survey we will mainly concentrate on acoustics in BECs,
and give some feel for where we stand and what the near-term prospects
are.  Acoustic analogues of black holes are formed by supersonic fluid
flow~\cite{Unruh,Visser}. The flow entrains sound waves and forms a
trapped region from which sound cannot escape.  The surface of no
return, the acoustic horizon, is qualitatively very similar to the
event horizon of a general relativity black hole; in particular
Hawking radiation (in this case a thermal bath of {\emph{phonons}}
with temperature proportional to the ``surface gravity'') is expected
to occur~\cite{Unruh,Visser}.  There are at least three physical
situations in which acoustic horizons are {\emph{known}} to occur:
Bondi--Hoyle accretion~\cite{Bondi-Hoyle}, the Parker
wind~\cite{Parker} (coronal outflow from a star), and supersonic wind
tunnels. Recent improvements in the creation and control of
Bose--Einstein condensates (see \eg,~\cite{Dalf99,Hau}) have lead to a
growing interest in these systems as experimental realizations of
acoustic analogs of event horizons.  In reference~\cite{Laval} we
considered supersonic flow of a BEC through a Laval nozzle
(converging-diverging nozzle) in a quasi-one-dimensional
approximation. We showed that this geometry allows the existence of a
fluid flow with acoustic horizons without requiring any special
external potential, and we then studied this flow with a view to
finding situations in which the Hawking effect is large.  We were able
to present simple physical estimates for the ``surface gravity'' and
Hawking temperature, and so to identify an experimentally plausible
configuration with a Hawking temperature of order $70$~n~K; this
figure should be contrasted with the critical condensation temperature
which is of the order of $90$~n~K. We stress that in present day
experiments the actual physical temperature of the condensate,
although difficult to measure, is believed to lie well below this
critical temperature.

\subsection{From Gross--Pitaevskii to hydrodynamics}

Bose--Einstein condensates are most usefully described by the
nonlinear Schr\"o\-dinger equation, also called the Gross--Pitaevskii
equation, or sometimes the time-depend\-ent Landau--Ginsburg equation:
\be
- i \hbar \; \partial_t \psi(t,\vec x) = - {\hbar^2\over2m}\nabla^2
\psi(t,\vec x) + \lambda \; ||\psi||^2 \; \psi(t,\vec x).
\ee
(We have suppressed the externally applied trapping potential for
algebraic simplicity. For many technical details, and various
extensions of the model, see {\Barcelo} \etal~\cite{Barcelo}.
That reference also contains an extensive background bibliography.)
Now use the Madelung representation~\cite{Madelung} to put the
Schr\"odinger equation in ``hydrodynamic'' form:
\be
\psi = \sqrt{\rho} \; \exp(-i\theta\; m/\hbar).
\ee
Take real and imaginary parts: The imaginary part is a continuity
equation for an irrotational fluid flow of velocity $\vec
v\equiv\nabla\theta$ and density $\rho$; while the real part is a
Hamilton--Jacobi equation (Bernoulli equation; its gradient leads to
the Euler equation). Specifically:
\be
\partial_t \rho + \nabla\cdot(\rho \; \nabla \theta) = 0.
\ee
\be
\frac{\partial}{\partial t} \theta + {1\over2}(\nabla \theta)^2 
+ {\lambda \; \rho \over m}
- {\hbar^2\over2m^2}\; 
{\Delta\sqrt\rho\over\sqrt\rho}= 0.
\ee
That is, the nonlinear Schr\"odinger equation is completely equivalent
to irrotational inviscid hydrodynamics with a particular form for the
enthalpy
\be
h = \int {\d p \over \rho} = {\lambda \; \rho\over m}, 
\ee
plus a peculiar derivative self-interaction:
\be
V_Q = - {\hbar^2\over2m^2}\; {\Delta\sqrt\rho\over\sqrt\rho}.
\ee
The equation of state for this ``quantum fluid'' is calculated from the
enthalpy
\be
p = {\lambda \;\rho^2\over2m}. 
\ee
The corresponding speed of sound is
\be
c_s^2=  {\d p\over\d\rho} 
= {\lambda \; \rho\over m}.
\ee

\subsection{Acoustic metric}

To now extract a Lorentzian geometry, linearize around some
background. In the low-momentum limit it is safe to neglect $V_Q$. It
is a by now standard result that the {\emph{phonon}} is a massless
minimally-coupled scalar that satisfies the d'Alembertian equation in
the effective (inverse) metric~\cite{Unruh,Visser,Barcelo}
\be
g^{\mu\nu}(t,\vec x) \equiv 
{\rho_0\over c_s}
\left[ \matrix{-1&\vdots&-v_0\cr
               \cdots\cdots&\cdot&\cdots\cdots\cdots\cdots\cr
	       -v_0&\vdots&(c_s^2 \;{\mathbf{I}} - v_0 \otimes v_0)\cr } 
\right].	       
\ee
Here
\be
c_s^2 \equiv {\lambda \; \;\rho_0\over m}; 
\qquad\qquad
v_0 = {\nabla \theta_0}.
\ee
It cannot be overemphasized that low-momentum phonon physics is
completely equivalent to (scalar) quantum field theory in curved
spacetime.  That is, everything that theorists have learned about
curved space QFT can be carried over to this acoustic system, and
conversely acoustic experiments can in principle be used to
experimentally investigate curved space QFT.  In particular, it is
expected that acoustic black holes (dumb holes) will form when the
condensate flow goes supersonic, and that they will emit a thermal
bath of Hawking radiation at a temperature related to the physical
acceleration of the condensate as it crosses the acoustic
horizon~\cite{Unruh,Visser,Barcelo,Garay}.  For completeness we
mention that the metric is
\be
g_{\mu\nu}(t,\vec x) 
\equiv \; {\rho_0\over c_s} \;
\left[ \matrix{-(c_s^2-v_0^2)&\vdots&-{\vec v}_0\cr
               \cdots\cdots\cdots\cdots&\cdot&\cdots\cdots\cr
	       -{\vec v}_0&\vdots& \mathbf{I}\cr } \right],
\ee
so the space-time interval can be written~\footnote{
{Although Eq.~\ref{eq:metr} seems to imply that a standard BEC can
only simulate metrics with conformally flat spatial sections, it can
nevertheless be shown that if the condensate is characterized by some
anisotropic mass tensor (realized, e.g., via some doping gradient)
then non-conformally flat spatial sections could also be
simulated~\cite{Barcelo}.}
}
\be
\d s^2 =
{\rho_0\over c_s} \;
\left[
- c_s^2 \; \d t^2 + ||\d\vec x - \vec v_0 \; \d t||^2 
\right].
\label{eq:metr}
\ee
The low-momentum phonon physics looks completely Lorentz
invariant. (This is an acoustic Lorentz invariance mind you, with the
speed of sound doing duty for the speed of light~\cite{Visser}.)

\subsection{Hawking effect}

\subsubsection{Laval nozzle} 

A general problem with the experimental construction of acoustic
horizons is that many of the background fluid flows so far studied
seem to require very special fine-tuned forms for the external
potential. (See \eg, the Schwarzschild-like geometry in
reference~\cite{Unexpected}.) In this respect a possible improvement
toward the realizability of acoustic horizons is the construction of a
flow in a trap which ``geometrically constrains'' the flow in such a
way as to replace the need for a special external potential. An
example of such a geometry is the so called Laval nozzle
(converging-diverging nozzle). In particular we shall consider a pair
of Laval nozzles; this provides a system which includes a region of
supersonic flow bounded between two subsonic regions.
\begin{figure}[htb]
\vbox{%
\hfil
\scalebox{1.00}{\includegraphics{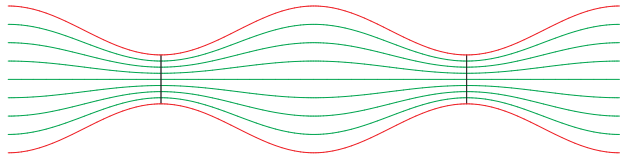}}
\hfil
}
\bigskip
\caption[Pair of Laval nozzles]{%
{\sl A pair of Laval nozzles: The second constriction is used to bring
the fluid flow back to subsonic velocities. 
\smallskip}} 
\end{figure}
Consider such a nozzle pointing along the $z$ axis.  Let the cross
sectional area be denoted $A(z)$.  We apply, with appropriate
modifications and simplifications, the calculations of
references~\cite{Laval} and~\cite{Unexpected}. The crucial
approximation is that transverse velocities (in the $x$ and $y$
directions) are small with respect to velocity along the $z$
axis. Then, assuming steady flow, we can write the continuity equation
in the form
\be
\rho(z) \; A(z) \; v(z) = J; \qquad J = \hbox{constant.}
\label{eq:cont}
\ee
The Euler equation (which we simplify by excluding external forces $\d
\Phi/\d z$, and excluding internal viscous friction $f_v$) reduces to
\be
\rho \; v \; {\d v\over \d z} = - {\d p\over\d z}.
\label{eq:eul}
\ee
Finally, we assume a barotropic equation of state $\rho = \rho(p)$,
and define $X' = \d X/\d z$.  Then continuity implies
\be
\rho' = - \rho \; {(A v)'\over(A v)} = 
-\rho \left[ {A'\over A} + {v'\over v}\right] = 
-\rho  \left[ {A'\over A} + {a\over v^2}\right],
\ee
while Euler implies
\be
\rho \;a = - {\d p\over\d\rho} \; \rho'.
\ee
Defining the speed of sound by $c^2 = \d p/\d\rho$, and eliminating
$\rho'$ between these two equations yields a form of the well-known
``nozzle equation''
\be
a =  - {v^2\; c^2\over c^2 - v^2} 
\left[ {A'\over A} \right].
\label{E:nozzle}
\ee
The presence of the factor $ c^2 - v^2$ in the denominator is crucial
and leads to several interesting physical effects. For instance, if
the physical acceleration is to be finite at the acoustic horizon, we
need
\be
A' \;\to\; 0.
\label{E:fine-tune}
\ee
This is a fine-tuning condition that forces the acoustic horizon
(technically, the acoustic ergosurface) to form at exactly the
narrowest part of the nozzle. (If external body forces and internal
friction are not neglected, then there is a precise relationship
between these forces and the location of the horizon.)  Experience
with wind tunnels has shown that the flow will indeed self-adjust (in
particular, the location of the acoustic horizon will self-adjust) so
as to satisfy this fine tuning. We can now calculate the
acceleration of the fluid at the acoustic horizon by adopting
L'\Hospital's rule. 
\be
a_H = {- c^4_H A''_H/A_H \over (c^2)'_H - 2 a_H }.
\ee
Now use the fact that
\bea
(c^2)' &\equiv& {\d^2p\over\d\rho^2} \; \rho' =   
-\rho\; {\d^2p\over\d\rho^2} \; {(A\;v)'\over A\;v} 
\to
- \rho_H  \; {\d^2p\over\d\rho^2} \bigg|_{H}\; 
\left[{a_H\over c_H^2} \right].
\label{E:c2'}
\eea
Therefore
\be
a_H^2 = {c^4 \; A''/A \over 2 +  \rho   (\d^2p/\d\rho^2)/c^2}\bigg|_{H}.
\ee
That is
\be
a_H = \pm {c^2\over\sqrt{2A}}\; 
\left.\sqrt{A''\over1 + {\rho\over2c^2} [\d^2p/\d\rho^2]}\right|_{H}.
\ee
Thus the physical acceleration of the fluid as it crosses an acoustic
horizon is tightly constrained in terms of the speed of sound, the
geometry of the horizon ($A_H$ and $A''_H$), plus some information
coming from the equation of state.

\subsubsection{``Surface gravity''}

It is more useful to consider the ``surface gravity'' defined by the
limit of the quantity~\cite{Visser}
\be
g = -{1\over2}{\d(c^2-v^2)\over \d z}
\ee
It is this combination $g$, rather than the physical acceleration of
the fluid $a$, that more closely tracks the general relativistic
notion of ``surface gravity'', and it is the limit of this quantity as
one approaches the acoustic horizon that enters into the Hawking
radiation calculation~\cite{Essential}.  Note that
\be
g =  a - {1\over2}(c^2)'.
\ee
This implies, in particular, that the fine-tuning (\ref{E:fine-tune})
used to keep $a$ finite at the acoustic horizon will also keep $g$
finite there. In particular
\be 
g_H = a_H \left[ 1 + {\rho\over2c^2} {\d^2 p\over\d\rho^2} \right]_H, 
\ee
and so
\be
g_H = 
\pm
{c^2_H \over \sqrt{2A_H}} \; 
\left.\sqrt{ 1 + {\rho\over2c^2} {\d^2 p\over \d\rho^2}}\right|_H \; 
\sqrt{A''_H}.
\label{E:g_H}
\ee 
The first factor is of order $c^2_H/R$, with $R$ the minimum radius of
the nozzle, while the second and third factors are square roots of
dimensionless numbers. This is in accord with our intuition based on
dimensional analysis~\cite{Visser,Unexpected}. If $A''<0$,
corresponding to a maximum of the cross section, then $a_H$ and $g_H$
are imaginary which means no event horizon can form there. The two
signs $\pm$ correspond to either speeding up and slowing down as you
cross the horizon, both of these must occur at a minimum of the cross
sectional area $A''>0$. (If the flow accelerates at the horizon this
is a black hole horizon [future horizon]; if the flow decelerates
there it is a white hole horizon [past horizon]. See Figure 1.) If the
nozzle has a circular cross section, then the quantity $A''_H$ is
related to the longitudinal radius of curvature $R_{c}$ at the throat
of the nozzle, in fact
\be
A_H'' = \pi {R \over R_c}.
\ee
%

\subsubsection{Bose--Einstein condensate}
The technological advantages provided by the use of BECs as a working
fluid for acoustic black holes have been discussed by
Garay~\etal~\cite{Garay} (see also reference~\cite{Barcelo} for a
discussion of plausible extensions to that model). The present
discussion can be interpreted as a somewhat different approach to the
same physical problem, side-stepping the technical complications of
the Bogoliubov equations in favour of a more fluid dynamical point of
view. For a standard BEC
\be
c^2 = {\lambda \rho\over m}.
\ee
Then
\be
\rho \left[{\d^2p\over\d\rho^2}\right] = \rho {\d(c^2)\over\d\rho} = c^2,
\ee
while
\be
1 + {1\over2}  {\rho\over c^2} \left[{\d^2p\over\d\rho^2}\right] = {3\over2}.
\ee
So we have, rather simply
\be
a_H = \pm{c^2_H\over\sqrt{A_H}}\; \sqrt{A''_H/3}.
\ee
Similarly
\be
g_H = \pm{c^2_H \over \sqrt{A_H}} \; \sqrt{3A''_H/4}.
\ee 
This implies, at a black hole horizon [future horizon], a Hawking
temperature~\cite{Unruh,Visser,Essential}
\be
k_{\rm B} T_H = {\hbar g_H\over2\pi c_H} = 
\hbar {c_H \over 2\pi \sqrt{A_H}} \; \sqrt{3A''_H\over4}.
\ee
Ignoring the issue of gray-body factors (they are a refinement on the
Hawking effect, not really an essential part of the physics), the
phonon spectrum peaks at
\be
\omega_{\mathrm{peak}} =  {c_H \over 2\pi \sqrt{A_H}} \; \sqrt{3A''_H\over4},
\ee
that is
\be
\lambda_{\mathrm{peak}} =4\pi^2 \sqrt{A_H} \; \sqrt{4\over3A''_H}.
\ee
This extremely simple result relates the typical wavelength of the
Hawking emission to the physical size of the constriction and a factor
depending on the flare-out at the narrowest point. Note that you
cannot permit $A''_H$ to become large, since then you would violate
the quasi-one-dimensional approximation for the fluid flow that we
have been using in this note. (There is of course nothing physically
wrong with violating the quasi-one-dimensional approximation, it just
means the analysis becomes more complicated. In particular, if there
is no external body force and the viscous forces are zero then by
slightly adapting the analysis of~\cite{Unexpected} the acoustic
horizon [more precisely the ergo-surface] is a minimal surface of zero
extrinsic curvature.)  The preceding argument suggests strongly that
the best we can realistically hope for is that the spectrum peaks at
wavelength
\be
\lambda_{\mathrm{peak}} \approx \sqrt{A_H}.
\ee
(Note that this is the analog, in the context of acoustic black holes,
of the fact that the Hawking flux from general relativity black holes
is expected to peak at wavelengths near the physical diameter of the
black hole, its Schwarzschild radius --- up to numerical factors
depending on charge and angular momentum.)  You can (in principle) try
to adjust the equation of state to make the second factor in
(\ref{E:g_H}) larger, but this is unlikely to be technologically
feasible.

\subsubsection{Physical estimates}
It is the fact that the peak wavelength of the Hawking radiation is of
order the physical dimensions of the system under consideration that
makes the effect so difficult to detect. This suggests that it might
be useful to look for indirect effects. In particular, in BECs it is
common to have a sound speed of order $6\;\hbox{mm/s}$.  If one then
chooses a nozzle of diameter about 1 micron, and a flare-out of
$A''_H\approx 1$, then $T_H \approx 7 \;\hbox{n K}$.  Compare this to
the condensation temperature required to form the BEC
\be
T_{\mathrm{condensate}} \approx 90 \;\hbox{n K}.
\ee
We see that in this situation the Hawking effect, although tiny, is at
least comparable in magnitude to other relevant temperature scales.
Moreover recent experiments indicate that it is likely that these
figures can be improved.  In particular, the scattering length for the
condensate can be tuned by making use of the so called Feshbach
resonance~\cite{Feshbach}. This effect can be used to increment the
scattering length; factors of up to 100 have been experimentally
obtained~\cite{Feshbach2}. Therefore the acoustic propagation speed,
which scales as the square root of the scattering length, could
thereby be enhanced by a factor up to 10. This suggests that it might
be experimentally possible to achieve $c_H\approx 6\;\hbox{cm/s}$, and
so
\be
T_H \approx 70\;\hbox{n K};
\ee
which places us much closer to the condensation temperature.  The
speed of sound can also be enhanced by increasing the density of the
condensate (propagation speed scales as the square root of the
density). In all of these situations there is a trade-off: For fixed
nozzle geometry the Hawking temperature scales as the speed of sound,
so larger sound speed gives a bigger effect but conversely makes it
more difficult to set up the supersonic flow.

The current analysis is purely ``hydrodynamic'', and does not seek to
deal with the ``quantum potential'' --- the fact that the dispersion
relation is at high momenta modified in such a way as to recover
``infinite'' propagation speed as in the Bogoliubov dispersion
relation~\cite{cpt01}.  This issue has relevance to the so-called
trans-Planckian problem (which in this BEC condensate context becomes
a trans-Bohrian problem).  Fortunately it is known, thanks to model
calculations in field theories with explicit high-momentum cutoffs,
that the low energy physics of the emitted radiation is largely
insensitive to the nature and specific features of the cutoff.

To summarize: this analysis complements that of Garay
\etal~\cite{Garay}, in that it provides a rationale for simple
physical estimates of the Hawking radiation temperature without having
to solve the full Bogoliubov equations. Additionally, the current
analysis provides simple numerical estimates of the size of the effect
and identifies several specific physical mechanisms by which the
Hawking temperature can be manipulated: via the speed of sound, the
nozzle radius, the equation of state, and the degree of flare-out at
the throat.

\subsection{Bogolubov dispersion relation}
\label{s:bogolubov}

However, there is a bit of a puzzle hiding in this analysis: We
started with the nonlinear Schr\"odinger equation. That equation is
parabolic, so we know that the characteristics move at infinite speed.
How did we get a hyperbolic d'Alembertian equation with a finite
propagation speed?  The subtlety resides in neglecting the
higher-derivative term $V_Q$. To see this, keep $V_Q$, and go to the
eikonal approximation. One obtains the dispersion
relation~\cite{Barcelo,cpt01}
\be
\left(\omega - \vec v_0 \cdot \vec k\right)^2 =
{c_s^2 \;k^2} +\left({\hbar\over2m} k^2\right)^2.
\ee
This is the curved-space generalization of the well-known
Bogolubov dispersion relation. Equivalently
\be
\omega=  \vec v_0 \cdot \vec k  +
\sqrt{ {c_s^2 \;k^2} +\left({\hbar\over2m} k^2\right)^2 }.
\ee
The group velocity is
\be
{\vec v}_g = {\partial\omega\over\partial\vec k} =  
\vec v_0  
+
{ \left(c_s^2+{\hbar^2 \over 2 m^2}k^2\right) 
\over 
\sqrt{c_s^2 k^2+\left({\hbar \over 2 m}\;k^2\right)^2} }
\; \vec k,
\ee
while for the phase velocity
\be
\vec v_p = {\omega\; \hat k\over||k||} =  (v_0 \cdot \hat k) \; \hat k  
+
\sqrt{c_s^2+{\hbar^2 \;k^2\over 4 m^2}} \;
\; \hat k.
\ee
Both group and phase velocities have the appropriate relativistic
limit at {\emph{low}} momentum, but then grow without bound at
{\emph{high}} momentum, leading to an infinite signal speed and the
recovery of the parabolic nature of the differential equation at high
momentum. ($k \gg k_c \equiv m\; c_s/\hbar$; equivalently in terms of
the acoustic Compton wavelength $\lambda \ll \lambda_C \equiv
\hbar/(m\; c_s)$.)

To investigate the situation a little more deeply, consider:
\be
\omega(k) = \sqrt{ m_0^2 + k^2 + \left({k^2\over2m_\infty}\right)^2 }.
\label{eq:gendisprel}
\ee
(This is equivalent to the original Bogolubov dispersion relation.  We
have set the background flow $v_0$ to zero.  In BEC condensates $m_0 =
0$ but there are other condensed matter systems where it need not be
zero. Additionally $c=\hbar=1$ for simplicity.)  Then at low momenta
($k \ll m_0$) the dispersion relation is Newtonian
\be
\omega(k) = m_0 + {k^2\over2m_0} + O(k^4).
\ee
while at intermediate momenta ($m_0 \ll k \ll m_\infty$) it is
(approximately) relativistic. Perhaps surprisingly at large momenta
($k \gg m_\infty$) the dispersion relation again takes on Newtonian
form
\be
\omega(k) = {k^2\over2m_\infty} + m_\infty + O(k^{-2}),
\ee
and explicitly deviates from Lorentz symmetry. (Even more complicated
deviations from Lorentz symmetry are possible, see for example
reference~\cite{re-entrant}.)

The implication is this: If we consider a mode that far away from the
horizon has a wave vector $k$ that is well inside the ``phonon''
region of the dispersion relation, and then follow that mode back
until it approaches the horizon, then near the horizon $k$ diverges
and the mode leaves the ``phonon'' region. It enters the ``particle''
region of the dispersion relation. This is the analog, in this
particular condensed matter context, of the so-called trans-Planckian
problem of general relativistic black hole physics. Fortunately it is
now realised that the low-$k$ far-from-the-horizon physics of the
Hawking effect is largely insensitive to the precise details of how
the dispersion relation is modified by high-$k$ near-horizon physics.
It is only part of the near-horizon physics, specifically the
``surface gravity'' that is really important in regards to the Hawking
effect. 

As a closing comment we would like to add that dispersion relation of
the form~(\ref{eq:gendisprel}), and with quadratic or cubic deviations
from Lorentz invariance, have also been encountered in several
approaches to quantum gravity (see e.g.~\cite{CPT01b}) and that there
have been recent attempts to test these ideas via astrophysical
observation. (See e.g.~\cite{sigl,ljm01} and references therein.)

\section{Slow light}
\setcounter{equation}{0}

Slow light systems, photon pulses with anomalously low group
velocities engendered by electromagnetically induced transparency
(EIT) in an otherwise opaque medium, have also been mooted as being
experimentally interesting avenues towards building analogue black
holes~\cite{Leonhardt,Visser2,Leonhardt2}.  One of the key issues here
is that EIT intrinsically requires one to work in a narrow frequency
range close to an atomic resonance; the resulting analogue black hole
will trap pulses of light only over a very narrow frequency range,
outside of which the medium is typically opaque.

Although the technology for building and manipulating slow light
systems is developing at an extremely rapid pace~\cite{Hau}, this
intrinsic limitation to working in a narrow frequency range somewhat
obscures the meaning of Hawking radiation and makes it less clear just
what signal should be looked for. For a discussion of the possibilities
see~\cite{Leonhardt3}.

\section{Quasiparticles}
\setcounter{equation}{0}

The use of superfluid quasiparticles, in particular the quasiparticles
and domain walls of liquid He$_3$A, has been investigated by Jacobson
and Volovik~\cite{Jacobson-Volovik}. A particularly nice feature is
due to the two-fluid nature of the system, in that in this system it
seems possible to arrange a wide separation between the Landau
critical velocity and the velocity relevant to defining the horizon at
which the Hawking phenomenon is expected to occur.

\section{Normal modes}
\setcounter{equation}{0}

The sheer number of different physical systems in which analogue models
for general relativity may be found is indicative of a deep underlying
principle. Indeed, finding an approximate Lorentzian geometry is
really just a matter of picking an arbitrary physical system,
isolating a particular degree of freedom that is approximately
decoupled from the rest of the physics, and doing a low-momentum
field-theory normal-modes analysis~\cite{normal-modes,emergent}.

Roughly speaking: in any hyperbolic system of differential equations
(no matter how derived) there are by definition wave-like
solutions~\cite{Courant,EDM}. The set of admissible wavevectors
associated with these wave-like solutions can be used to define
(modulo some nasty complications we defer to the technical
literature~\cite{normal-modes2}) a cone-like structure in momentum
space, and hence a conformal class of Lorentzian-signature metrics.
For this reason the emergence of Lorentzian-signature effective
metrics is an almost generic aspect of low-momentum physics.

\section{Emergent gravity}
\setcounter{equation}{0}

So far, the entire discussion has been about models {\emph{of}}
gravity, models that reproduce the {\emph{kinematics}}. If one wants
to make a bolder proposal, that analogue models might be useful for
generating models {\emph{for}} gravity, models that reproduce the
Einstein--Hilbert {\emph{dynamics}} (or some approximation thereto),
then the situation is considerably more subtle and
tentative. Kinematics is relatively easy, and is in some sense
generic. Einstein--Hilbert dynamics is trickier --- to get an
``emergent gravity'' arising from these analogue models will require
some variant of Sakharov's notion of ``induced
gravity''~\cite{Sakharov}.  A useful observation in this regard is
that any curved-space relativistic quantum field theory will
automatically generate an Einstein--Hilbert counterterm through
one-loop effects~\cite{emergent}. In heat kernel language, the first
Seeley--DeWitt coefficient generically contains a term proportional to
the Einstein--Hilbert action, and after renormalization this
generically provides an Einstein--Hilbert term in the effective
action~\cite{zeta}. Unfortunately the same logic provides an
uncontrolled cosmological constant from the zeroth Seeley--DeWitt
coefficient, plus quadratic curvature-squared terms from the second
Seeley--DeWitt coefficient, so the argument is not fully
acceptable. Furthermore there are technical issues involved in
specifying the volume of the function space on which this effective
action is defined. To get Einstein gravity one needs both an
Einstein--Hilbert action and the freedom to perform arbitrary metric
variations. Though the situation is still far from clear, interest in
these possibilities is both long-standing (see for instance the
sub-manifold models in references~\cite{Regge,Deser}) and
ongoing~\cite{emergent,normal-modes2,Chapline,Volovik-Report}.

\section{Discussion}
\setcounter{equation}{0}

In this mini-survey we have seen how an effective metric emerges as a
low-energy low-momentum approximation in certain physical
systems. Indeed we have been able to argue that the emergence of such
effective metrics is an almost generic consequence of performing a
``normal modes'' analysis on an arbitrary field theory. Once one has
an effective metric in hand (no matter how derived), all kinematic
aspects of general relativity can in principle be carried over to
these analogue systems --- in particular all curved-space field theory
(both classical and quantum) finds a natural home in these systems.

The most stunning feature of these analogue models is the ability to
generate analogue horizons (analogue black holes) and more
specifically, the possibility of detecting an analogue form of Hawking
radiation. In the body of this article we have specifically considered
the use of acoustics in Bose--Einstein condensates as a particularly
promising analogue system. This particular model stands out for purely
technological reasons --- the condensation temperature, required to
form the condensate in the first place is of order $90\;\hbox{nK}$,
which is considerably less than 1 order of magnitude away from the
estimates of the relevant Hawking temperature. It is this congruence
between two important physical scales that makes this particular
system so interesting. Many condensed matter systems are capable of
mimicking curved space quantum field theory; this particular condensed
matter system does so in a particularly interesting manner that seems
amenable to experimental probes in the not too distant future.

\section*{Acknowledgements}
Matt Visser was supported by the US DOE. Stefano Liberati was
supported by the US NSF. Carlos {\Barcelo} was supported by the Spanish
MCYT, and is now supported by a European Community Marie Curie grant.


\end{document}